\documentclass[pre,onecolumn,apssuperscriptaddress,longbibliography]{revtex4-1}
\usepackage{amsmath,amssymb,graphicx,cases}
\usepackage{epsfig}
\usepackage{textcomp}
\usepackage{epstopdf}
\usepackage{float}
\usepackage{braket}
\usepackage[normalem]{ulem}
\usepackage{enumerate}
\usepackage{enumitem}

\usepackage{ae,aecompl}
\usepackage{lmodern}

\usepackage{color}
\definecolor{darkblue}{rgb}{0,0,0.6}
\definecolor{darkred}{rgb}{0.6,0,0}
\definecolor{darkgrey}{rgb}{0.6,0.6,0.6}

\newcommand{\stkout}[1]{\ifmmode\text{\sout{\ensuremath{#1}}}\else\sout{#1}\fi}

\usepackage[colorlinks=true, urlcolor=blue, anchorcolor=blue, citecolor=blue,filecolor=blue,linkcolor=blue,menucolor=blue]{hyperref}

\begin{document}

	\title{Exact fluctuating hydrodynamics of active lattice gases --  \\
	Typical fluctuations}

\author{Tal Agranov}
\affiliation{Department of Physics,
	Technion-Israel Institute of Technology,
	Haifa, 3200003, Israel.}
\affiliation{Department of Applied Mathematics and Theoretical Physics,
	University of Cambridge, 
	Wilberforce Road, Cambridge CB3 0WA, United Kingdom.}
\author{Sunghan Ro}
\affiliation{Department of Physics,
	Technion-Israel Institute of Technology,
	Haifa, 3200003, Israel.}
\author{Yariv Kafri}
\affiliation{Department of Physics,
	Technion-Israel Institute of Technology,
	Haifa, 3200003, Israel.}
\author{Vivien Lecomte}
\affiliation{Université Grenoble Alpes,
	CNRS, LIPhy, 
	38000 Grenoble, France.}

	\begin{abstract}
We extend recent results on the exact hydrodynamics of a system of diffusive active particles displaying a motility-induced phase separation to account for typical fluctuations of the dynamical fields. By calculating correlation functions exactly in the homogeneous phase, we find that two macroscopic length scales develop in the system. The first is related to the diffusive length of the particles and the other to the collective behavior of the particles. The latter diverges as the critical point is approached. 
Our results show that the critical behavior of the model in one dimension belongs to the universality class of a mean-field Ising model, both for static and dynamic properties, when the thermodynamic limit is taken in a specified manner. 
The results are compared to the critical behavior exhibited by the ABC model. In particular, we find that in contrast to the ABC model the density large deviation function, at its Gaussian approximation, does not contain algebraically decaying interactions but is of a finite, macroscopic, extent which is dictated by the diverging correlation length.
	\end{abstract}	
\maketitle

\tableofcontents

\section{Introduction}
In recent years there has been a lot of interest in the statistical mechanics of active systems. These consist of particles that self-propel by consuming energy. They have attracted much attention both due to their relevance to biological and synthetic materials, and due to the host of novel non-equilibrium behaviors that arise from the interplay between the interactions of the agents and the non-equilibrium drive that acts on the single-particle level~\cite{ramaswamy_mechanics_2010,thompson_lattice_2011,romanczuk_active_2012,marchetti_hydrodynamics_2013,takatori_towards_2015,ramaswamy_active_2017,needleman_active_2017,loi_effective_2008,solon_pressure_2015,fodor_how_2016,soto2014run,slowman2016jamming}. Most theoretical studies of active matter involve either phenomenological approaches based on field theoretical models constructed using symmetry arguments, or derivations which start from microscopic dynamics and involve mean-field approximations and gradient expansions, see for example~\cite{wittkowski2014scalar,nardini2017entropy,solon_generalized_2018,solon_generalized_2018-1,tjhung_cluster_2018,peshkov2012nonlinear,bertin2013mesoscopic,bertin2015comparison}. While these have been extremely successful in predicting and explaining a large variety of interesting behaviors, there is a clear lack of exactly solvable models in the field. Technically, it is the presence of interactions and the non-equilibrium drive that make active systems very difficult to study using exact methods. Indeed, even the full dynamical problem of two active particles on a lattice interacting via hard-core interactions is not trivial~\cite{mallmin2019exact}.

In parallel to the advancements in active matter, there have been major developments in our understanding of the fluctuating hydrodynamics of diffusive non-equilibrium systems. This started with the identification of the hydrodynamic equations which capture typical fluctuations by Spohn for a class of lattice models (see~\cite{spohn_large_1991} for a review) and the discovery of the generic long-range nature of correlations
when driven out of equilibrium, say by coupling the system to boundaries at different densities~\cite{spohn_long_1983}. 
More recently a formalism, known as the Macroscopic Fluctuation Theory (MFT), was put forward. The MFT utilizes fluctuating hydrodynamics equations, whose exact form can be derived in several cases starting from a microscopic description~\cite{derrida_non-equilibrium_2007,derrida_microscopic_2011,bertini_macroscopic_2015}.  It enables one to study the probability density and dynamics of large non-equilibrium fluctuations in spatially extended systems of diffusive interacting particles (see~\cite{bertini_macroscopic_2015} for a review) in the limit of a large number of degrees of freedom. In this limit, the probability distributions take an exponential form where the exponent (the `quasi-potential') is the direct analog of the equilibrium free energy. Using MFT, it was understood that for many non-equilibrium systems the probability density is highly non-trivial. For example, the probability distribution of the density field is generically a non-local functional of the field and the probability distributions of various quantities are often singular~\cite{bertini_macroscopic_2015,bodineau_current_2010,PhysRevLett.116.240603,PhysRevLett.118.030604,PhysRevLett.114.060601,Lecomte_2012,PhysRevE.99.052102,baek2015singularities}. To date, an analogous set of results for active matter systems does not exist. 

A major progress in this direction has recently been achieved in~\cite{erignoux_hydrodynamic_2018,kourbane-houssene_exact_2018}, where the \textit{noiseless} coarse-grained hydrodynamics of two active matter models, defined at the microscopic level, were derived. Here we focus on one of the models which on a microscopic scale, and in one dimension, is defined on a lattice where each site can be either empty or occupied by a right or a left moving particle. Right (left) moving particles perform a biased diffusion with a bias to the right (left) and a right moving particle can tumble into a left moving one and vice versa (see Fig.~\ref{schem} for an illustration). The model bears many similarities with the well-studied diffusive lattice gases~\cite{kipnis_scaling_1999,bertini_stochastic_2007,liggett_interacting_0000}. In particular, the motion of the particle is biased with a rate that scales as $L^{-1}$ with $L$ the system size (i.e.~the number of lattice sites), and the tumbling rates scale as $L^{-2}$.  This choice ensures a non-trivial diffusive scaling in the hydrodynamic limit (see details below). It is important to stress that this scaling of the rates places the model in a different class than active systems where the rates are not scaled with the system size. In particular, in contrast to the latter models where the only slow degree of freedom is the density, here both the polarization (defined as the density difference between right and left moves) and the density are slow degrees of freedom. Building on recent developments in the derivation of hydrodynamic equations in non-equilibrium systems~\cite{erignoux_hydrodynamic_2018}, the authors present the exact diffusive hydrodynamic description of the model~\cite{erignoux_hydrodynamic_2018,kourbane-houssene_exact_2018}. Interestingly, the model displays a motility-induced phase separation (MIPS), one of the most studied collective behavior in active matter. This corresponds to the ability of active systems to phase separate even when the interaction between the particles is purely repulsive~\cite{tailleur_statistical_2008,fily_athermal_2012,buttinoni_dynamical_2013,cates_when_2013,stenhammar_continuum_2013,redner_structure_2013,cates_motility-induced_2015,solon_active_2015,redner_classical_2016,whitelam_phase_2018,tjhung_cluster_2018,paliwal_chemical_2018,geyer_freezing_2019,solon_generalized_2018,solon_generalized_2018-1}. Using the hydrodynamic equations, the exact phase diagram was derived using the method presented in~\cite{solon_generalized_2018,solon_generalized_2018-1}. 

In this article, we build upon this work and complement the deterministic hydrodynamic equations with noise terms at the Gaussian level which captures typical fluctuations. This allows us to obtain two-point
correlation functions exactly to leading order in $1/L$,
and in regions of the phase diagram where the system is not phase separated.
We find that in these regions of the phase diagram the model exhibits two length scales each scaling with the system size. The first corresponds to the length scale associated with the typical distance diffused by the particle until it tumbles. The second describes the emergence of collective behavior in the system and diverges as the critical point is approached. Importantly, we characterize the critical behavior exhibited by the model exactly and we find that when the thermodynamic limit is taken in a specified way it is in the {\it Ising mean-field} universality class. While, as we explain below, we are not able to study the model exactly at the critical point, we find that the dynamics exhibit critical slowing down close to the critical point on scales smaller than the correlation length, with a mean-field model B exponent $z=4$. 

It is interesting to place these results in the context of recent studies on the universality of MIPS. This has been a topic of debate due to conflicting numerical and theoretical results. Studies based on a phenomenological theory for active systems, known as active model B+~\cite{wittkowski2014scalar,nardini2017entropy,tjhung_cluster_2018} suggest that MIPS belongs to the Ising universality class~\cite{Caballero_2018} (see also \cite{PhysRevLett.123.068002}). In contrast, several numerical studies present contradictory findings with some numerics suggesting an Ising universality~\cite{PhysRevLett.123.068002,maggi2021universality}, while others~\cite{PhysRevE.98.030601,dittrich2021critical} a different one. 
Note that, as stressed above, while these models are similar to the model we study, they belong to a distinct class since their transition rates do not scale with the system size $L$.
In particular, the standard active models do not phase separate in one dimension~\cite{thompson_lattice_2011}. 

Our results also allow us to consider the structure of the quasi-potential for the probability of profiles, at the Gaussian level. It was previously shown that lattice models which phase separate have a quasi-potential with non-local interactions. Here, in contrast, we find that the quasi-potential has a {\it local macroscopic structure} whose range grows as the critical point is approached. This, in turn, allows the system to supports a phase-separated state in one dimension. 

The outline of the paper is the following: in Sec.~\ref{sec:model-its-known}, we define the model and recall for completeness its (noiseless) hydrodynamics, which was derived in Ref.~\cite{kourbane-houssene_exact_2018}.
In Sec.~\ref{sec:fluct-hydr-active}, we describe our results, a set of fluctuating hydrodynamic equations with Gaussian noise, and the analytically computed two-point correlation functions for the different fields. We check the self-consistency of the calculation using a Ginzburg criterion and compare the results to numerical simulations of the lattice model. The quasi-potential (or free energy) is then computed at the quadratic level and its local structure is discussed.
In Sec.~\ref{sec:deriv-fluct-hydr}, we explain how to derive the Gaussian fluctuating hydrodynamic equations.
The derivation of the two-point correlation function is presented in Sec.~\ref{sec:derivation-two-point}.
We first derive the correlation functions in the limit of a small diffusive length of the active particles, and discuss the dynamic scaling exponent in Sec.~\ref{sec:dynamic}.
In Sec.~\ref{sec:finite-size-results}, we generalize the results to any diffusive length, which allows us to consider finite-size effects. In addition, it allows us to compare the observed transition to the one in the ABC model~\cite{evans_phase_1998,evans_phase_1998-1} --~another diffusive model which is known to exhibit phase separation in one dimension.
Finally, in Sec.~\ref{sec:discussion}, we conclude.

\section{The model and its known hydrodynamics}
\label{sec:model-its-known}

\begin{figure} [t!]
	\includegraphics[width=1.0\linewidth]{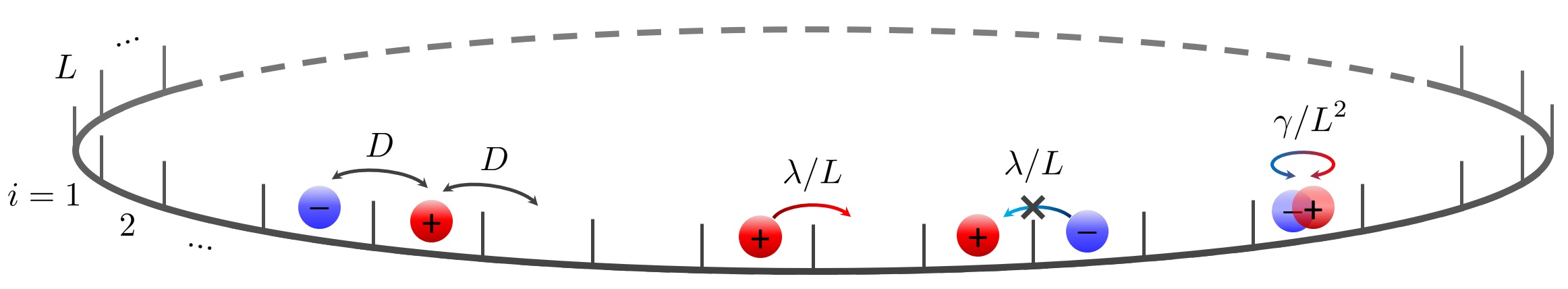}
	\caption{Schematic representation of the microscopic dynamics described in the text.}
	\label{schem}
\end{figure}

Following~\cite{kourbane-houssene_exact_2018}, we consider a one-dimensional lattice model constituted of $L$ sites and $N$ particles with a mean density $\rho_0 \equiv N/L$.
Each site $i$ can either be occupied by a~$+$~particle ($\sigma_i^+ = 1$ and $\sigma_i^- = 0$), a~$-$~particle ($\sigma_i^+ = 0$ and $\sigma_i^- = 1$), or be empty ($\sigma_i^+ = 0$ and $\sigma_i^- = 0$). 
The dynamics of the model is defined through the following independent processes (see Fig.~\ref{schem}):
\begin{enumerate}[label=(\roman*)]
	\item Diffusion: A pair of neighboring sites exchange their state with a rate $D$.
	\item Drift: A $+$ ($-$) particle jumps to the right (left) neighboring site with a rate $\lambda/L$, provided the target site is empty.
	\item Tumbling: A $+$ ($-$) particle tumbles into a $-$ ($+$) particle with a rate $\gamma/L^2$.
\end{enumerate}
The scaling of the rates with $L$ ensures that, in the hydrodynamic limit ($L\gg 1$), all processes occur on diffusive time scales. Indeed, the time it takes for particles to traverse a macroscopic system of size $L$ either via diffusive motion, or on account of self-propulsion, scales as $L^2$ which is also the time scale for tumbling events. 

Using the diffusive scaling $t\rightarrow t/L^2$ and $x=i/L$~\cite{kourbane-houssene_exact_2018}, so that $0\leq x \leq 1$, it is natural to define the coarse-grained fields $\rho_+ (x,t)$ and $\rho_- (x, t)$ in the large $L$ limit through
\begin{equation}
	\rho_{\pm}(x,t)=\frac{1}{2L^{\delta}}\sum_{|i-Lx|<L^{\delta}}\sigma_i^{\pm} \:,
\end{equation}
where the exponent $\delta$ can take any value in the range  $0<\delta<1$.  The resulting hydrodynamic equations of the model were shown in Ref.~\cite{kourbane-houssene_exact_2018} to be 
\begin{eqnarray}
	\partial_t\rho_+&=&D\partial^2_x\rho_+-\lambda\partial_x\left[\rho_+(1-\rho)\right]+\gamma(\rho_--\rho_+)\;,\label{eq:+}\\
	\partial_t\rho_-&=&D\partial^2_x\rho_-+\lambda\partial_x\left[\rho_-(1-\rho)\right]+\gamma(\rho_+-\rho_-)\:,\label{eq:-}
\end{eqnarray}
where $\rho(x, t) = \rho_+ (x,t) + \rho_- (x,t)$ is the total density of particles. 

For large enough drift $\lambda$ and mean density $\rho_0$, the hydrodynamic equations allow for non-homogeneous solutions. These obey the time-independent equations \begin{equation}\label{sta}
\left(\ell\partial_x\right)^2\rho_{\pm}\mp\text{Pe}\left(\ell\partial_x\right)\left[\rho_{\pm}\left(1-\rho\right)\right]\pm\left(\rho_--\rho_+\right)=0.
\end{equation}
Here $\ell \equiv \sqrt{D/\gamma}$ is the diffusive length  which prescribes the typical distance traveled by a particle only using diffusive steps before it tumbles. The P\'eclet number $\text{Pe}=\lambda/\sqrt{\gamma D}$ compares the persistence length $\lambda/\gamma$ to the diffusive one. Eq.~\eqref{sta} predicts that the length scale associated with spatially modulated solutions at fixed $\text{Pe}$ is controlled by $\ell$.
When this scale is very small compared to the macroscopic system size ($ \ell \ll 1$),
these solutions consist of sharply phase-separated regions with high- and low-density coexisting phases. The width of the domain walls separating them is set by $\ell$. The densities in the phases were calculated using an effective common tangent construction, and are independent of the mean density $\rho_0$, see \cite{solon_generalized_2018,solon_generalized_2018-1,kourbane-houssene_exact_2018}.
This phenomenon occurs only in the $\ell\rightarrow0$ limit and corresponds to MIPS.  

\begin{figure} [t!]
	\includegraphics[width=0.5\linewidth]{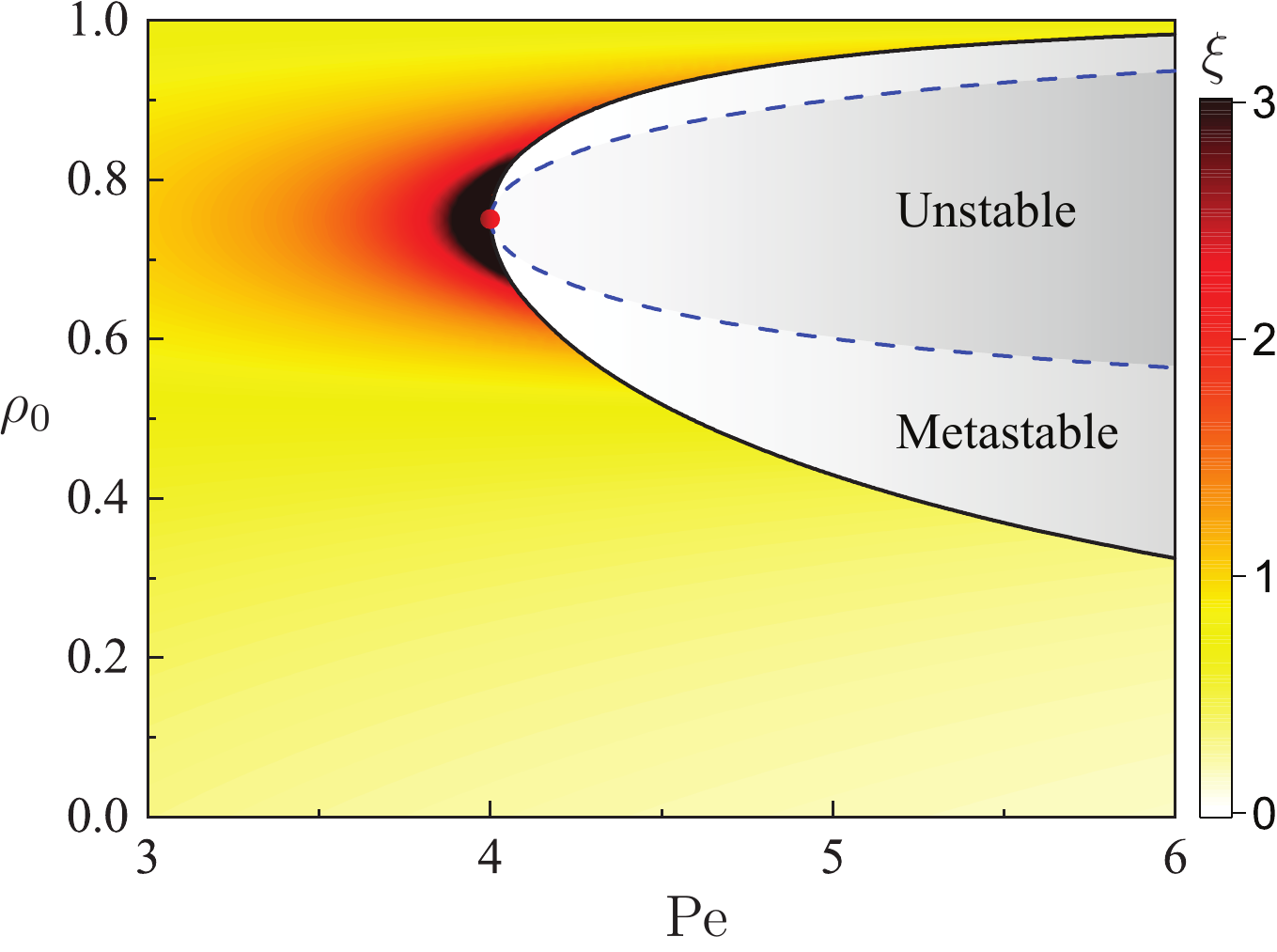}
	\caption{The phase diagram in the total mean-density $\rho_0$ and $\text{Pe}=\lambda/\sqrt{\gamma D}$ plane. The binodal curve is marked as a black solid line, while the blue dashed line indicates the spinodal line~$2-\text{Pe}^2(1-\rho_0)(2\rho_0-1)=0$. Both lines meet at the critical point $(\rho_0=3/4,\text{Pe}=4)$ which is marked by a red dot. 
The stationary configurations are phase-separated in the gray region, and otherwise homogeneous in the colored region.
The color code, given by the bar in the legend, indicates the value of the correlation length~\eqref{xi} derived using the fluctuating hydrodynamics.
 The binodal is obtained using the methods described in Ref.~\cite{solon_generalized_2018,solon_generalized_2018-1,kourbane-houssene_exact_2018} in the limit of small $\ell$.
}
	\label{fig:bin}
\end{figure}

The binodal curve has been calculated in~\cite{kourbane-houssene_exact_2018} in the small $\ell$ limit, and is shown in Fig.~\ref{fig:bin} for completeness. The figure also shows the spinodal line, beyond which the homogeneous state becomes linearly unstable. The spinodal line is given by $2-\text{Pe}^2(1-\rho_0)(2\rho_0-1)=0$. The critical point of the model 
$(\rho_c,\text{Pe}_c)=(3/4,4)$
is
located at the point where binodal and spinodal lines meet. 

The hydrodynamic equations presented above neglect the noise terms which on general grounds should scale as $L^{-1/2}$. The aim of this paper is to derive these at the Gaussian level and use them to compute two-point correlation functions. Of particular interest is their behavior near the critical point of the model. In what follows, before turning to the derivation of the different results, we summarize our main findings and contrast the behavior of the system with previous studies on other systems that phase separate in one dimension. For clarity, we first focus on the regime $\ell \ll 1$ (the case of finite $\ell$ is discussed in Sec.~\ref{sec:finite-size-results}).

\section{Fluctuating hydrodynamics of active lattice gas}
\label{sec:fluct-hydr-active}

\subsection{Complementing the hydrodynamic equations with a small Gaussian noise}
\label{sec:compl-hydr-equat}
While the hydrodynamic equations are exact in the $L \to \infty$ limit, any finite value of $L$ will lead to a fluctuating noisy dynamics. 
Our first result complements the hydrodynamic equations of~\cite{kourbane-houssene_exact_2018} with Gaussian noise terms which account for the \textit{typical} fluctuations $\delta\rho\sim L^{-1/2}$ around the most probable profile predicted by the deterministic equations~(\ref{eq:+})-(\ref{eq:-}). 
Adopting a representation in terms of the total particle density and polarization fields $\rho=\rho_++\rho_-$, $m=\rho_+-\rho_-$ we find 
that the fluctuating hydrodynamical equations take the form:
\begin{eqnarray}
\partial_t \rho&=& D \partial^2_{x}\rho-\lambda\partial_x[m(1-\rho)]+\frac{1}{\sqrt{L}} \sqrt{D}\,\partial_x\eta_\rho\label{eq:rho}\\
\partial_t m&=& D \partial^2_{x}m-\lambda\partial_x[\rho(1-\rho)]-2\gamma m+\frac{1}{\sqrt{L}}\Big( \sqrt{D} \,\partial_x\eta_m+2\sqrt{\gamma}\,\eta_K \Big) \:. \label{eq:m}
\end{eqnarray}
Here, $\eta_{\rho},\eta_{m}$, and $\eta_{K}$ are Gaussian white noises of zero mean and correlations given by
\begin{eqnarray}
\langle\eta_{\rho} (x,t)\eta_{\rho} (x',t')\rangle&=&2\rho (1-\rho)\,\delta (x-x')\delta (t-t')\quad;\quad\langle\eta_{\rho} (x,t)\eta_{m} (x',t')\rangle=2m (1-\rho)\,\delta (x-x')\delta (t-t'),
\label{covrho}\\
\langle\eta_{m} (x,t)\eta_{m} (x',t')\rangle&=&2 (\rho-m^2)\delta (x-x')\delta (t-t')\quad;\quad \langle\eta_K (x,t)\eta_K (x',t')\rangle=\rho\,\delta (x-x')\delta (t-t'),
\label{covm} \\
\langle\eta_{m} (x,t)\eta_{K} (x',t')\rangle&=&\langle\eta_{\rho} (x,t)\eta_{K} (x',t')\rangle= 0 \;. 
\label{eq:cov0}
\end{eqnarray}
The angular brackets denote an average over histories of the microscopic dynamics described in Sec.~\ref{sec:model-its-known}.
The noise terms are multiplicative, and as is usual in fluctuating hydrodynamics equations, it is straightforward to check that the scaling with $L^{-1/2}$ implies that the choice of time discretization, say It\=o or Stratonovich, is not important.
As we describe in Sec.~\ref{sec:deriv-fluct-hydr}, the equations are obtained through a superposition of known results for the ABC lattice gas~\cite{evans_phase_1998,evans_phase_1998-1} and a reaction process accounting for the tumbling.
It is straightforward to check that by rescaling $x$ by $\ell$ and $t$ with the timescale $1/\gamma$ (after the diffusive rescaling) the noiseless hydrodynamic part is entirely controlled by the P\'eclet number $\text{Pe}={\lambda}/{\sqrt{\gamma D}}$~\cite{kourbane-houssene_exact_2018}. The noise amplitudes in Eqs.~(\ref{eq:rho})-(\ref{eq:m}) then become proportional to $1/\sqrt{\ell L}$, indicating that the noise term keeps track of the lengthscale $\ell$ and the system size $L$.

\subsection{Two-point correlation functions in the homogeneous phase}
\label{sec:two-point-corr}

\begin{figure} [t!]
	\includegraphics[width=1.0\linewidth]{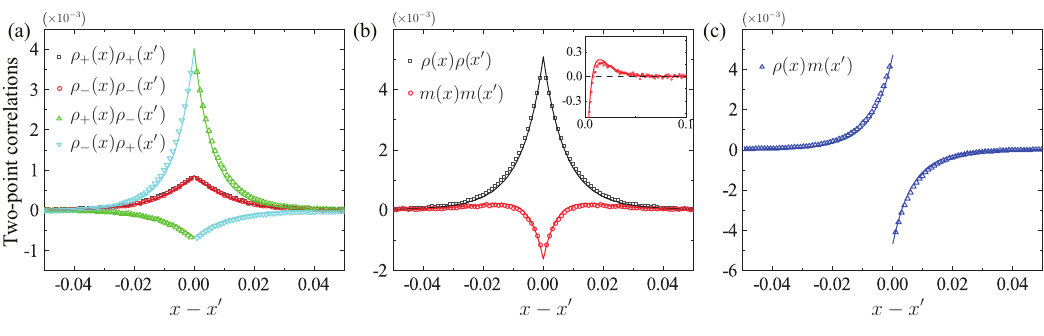}
	\caption{Two-point correlation functions of (a) type-specific density fields, (b) $\rho$ and $m$, and (c) cross-correlation between $\rho$ and $m$ are presented. Symbols show values obtained from simulation and lines correspond to analytic results obtained by using Eqs.~(\ref{eq:corr})-(\ref{eq:cross}). 
The inset of (b) illustrates the non-monotonous behavior of the correlation function $C_2^m$.
 The parameters used in the numerics are: $L=10^3$, $\rho_0 = 0.25$, $\gamma = 10^4$, $\mathrm{Pe} = 2$, and $D = 1$. }
	\label{Fig:twopoint}
\end{figure}

At large $L$, the hydrodynamic equations augmented with the Gaussian noise terms Eqs.~(\ref{eq:rho})-(\ref{eq:cov0}) are enough to give the exact two-point functions of the model~\cite{derrida_non-equilibrium_2007,PhysRevE.86.031106}
at leading order in $L^{-1}$. As detailed in Sec.~\ref{sec:derivation-two-point}, we find to this order that when the system is homogeneous, namely outside the binodal, the static connected two-point correlation functions are given by:
\begin{eqnarray}
&&
C_2^{\rho}
\equiv
\langle \delta\rho(x)\delta\rho(x')\rangle
=
 \frac{\rho_0(1-\rho_0)}{L}\left[\delta(x-x')-1\right]+
              \frac{ \text{Pe}^2\xi^2\rho_0^2(1-\rho_0)^2}
              {2\ell L(1-\xi^2)}
              \!\left[\!\left(e^{-\frac{1}{\ell}|x-x'|} -\ell\right)+\frac{1-2\xi^2}{\xi}\left(e^{-\frac{1}{\ell}\frac{|x-x'|}{\xi}}-\ell\xi\right)\!\right]~
\label{eq:corr}
\\
&&
C^m_2
\equiv
\langle\delta m(x)\delta m(x')\rangle
=
\frac{\rho_0}{L}\left[\delta(x-x')-1\right]+
			\frac{ \text{Pe}^2\xi^2\rho_0^2(1-\rho_0)(1-2\rho_0)}
			{2\ell L(1-\xi^2)}
			\left(e^{-\frac{1}{\ell}|x-x'|}-\frac{1}{\xi}e^{-\frac{1}{\ell}\frac{|x-x'|}{\xi}}\right), 
\label{Eq:c2_m}
\\
&&
C^{\rho,m}_2
\equiv
\left<\delta\rho (x)\delta m(x')\right>
=
\text{sign}(x-x')
\frac{ \text{Pe}\,\xi^2\rho_0^2(1-\rho_0)}{2\ell L(1-\xi^2)}
\left(e^{-\frac{1}{\ell}|x-x'|}+\frac{1-2\xi^2}{\xi^2}e^{-\frac{1}{\ell}\frac{|x-x'|}{\xi}}\right),\label{eq:cross}
\end{eqnarray}
where $\rho_0$ is the mean particle density, $\delta \rho(x)=\rho(x)-\rho_0$, $\delta m(x)=m(x)$ with the mean-magnetization zero, and the (dimensionless) correlation length $\xi$ is given by
\begin{equation}
\xi=\frac{1}{\sqrt{2-\text{Pe}^2(1-\rho_0)(2\rho_0-1)}}\:.\label{xi}
\end{equation}
The expressions are in excellent agreement with the Monte Carlo simulation results of the underlying microscopic dynamics of the lattice model, as illustrated in Fig.~\ref{Fig:twopoint}(b) and~\ref{Fig:twopoint}(c)~\footnote{%
For $\xi=1$, the expressions~\unexpanded{\eqref{eq:corr}-\eqref{eq:cross}} seems singular, due to the vanishing denominator $\sim 1-\xi$. Note however that the numerator vanishes at the same order in this limit, and the continuation in $\xi=1$ is analytical. E.g.,
\unexpanded{
\begin{equation}
\frac{1}{1-\xi^2}\Big(e^{-\frac{1}{\ell}|x-x'|}+\frac{1-2\xi^2}{\xi}e^{-\frac{1}{\ell}\frac{|x-x'|}{\xi}}\Big)
\xrightarrow[\xi=1]{}
\frac{3+\frac{1}{\ell}\,|x-x'|}{2}e^{-\frac{1}{\ell}|x-x'|}\;.
\end{equation}
}}.
The results~(\ref{eq:corr})-(\ref{eq:cross}) are valid for $\ell \ll 1$ and we give the expression for arbitrary $\ell$ in Sec.~\ref{sec:finite-size-results}. Several comments are in order. First, note that by setting the self-propulsion to zero by taking $\text{Pe}=0$, we are left only with local correlations as expected --~in this limit the model becomes identical to the simple symmetric exclusion model (SSEP) on a ring (see~\cite{derrida_microscopic_2011} for a review). Second, the integral of these expressions over each of the coordinates $x,x'$ separately, vanishes to leading order in $\ell$, as it should, due to total particle number conservation.  Finally, when $\ell=0$, so that the diffusive length is microscopic, one is left with local contributions only, encoded in the delta function.

When $\text{Pe}\neq0$, the model develops {\it large-scale correlations} of the order of the system size (recall that length scales are rescaled by the system size $L$). Indeed, in the hydrodynamic limit, any finite correlation length that does not scale with the system size will vanish and only manifest itself through the Dirac delta function term in the correlation function. 
It is well known that long-range correlations can appear in lattice gas systems when driven out of equilibrium, for example, by coupling the system to reservoirs of different densities at their boundaries~\cite{spohn_long_1983}. The large-scale correlations that we observe here are, in contrast, a consequence of the active dynamics of the particles. 

Interestingly there are two distinct length scales appearing in the correlation functions. One, $\ell$, is independent of the density and stems from the diffusive length of the particles. Recall, that here this length scale, which also controls the domain wall width in the phase-separated state, is assumed to be very small. Thus, the terms in the two-point functions associated with this scale are narrowly localized at $x=x'$. The second one, $\xi \ell$, is related to the collective behavior leading to MIPS. In particular, $\xi$ diverges upon approaching the critical MIPS point $(\rho_0=3/4,\text{Pe}=4)$ with a singular behavior of the form
\begin{equation}
\xi(\rho_0=3/4,\text{Pe})=\frac{1}{\sqrt{4-\text{Pe}}}+\mathcal O\big[(4-\text{Pe})^{1/2}\big],\label{div}
\end{equation}
corresponding to a mean-field like critical exponent $1/2$. Thus, when approaching the critical point, the length scale $\xi\ell$ approaches the scale of the system, and the associated terms in the expressions of the correlation functions span the entire system. As detailed below, at this stage our small $\ell$ approximation fails, and one has to resort to the finite $\ell$ expressions presented in Sec.~\ref{sec:finite-size-results}. Note also that the amplitude of the two-point correlation function~\eqref{eq:corr} diverges upon approaching the critical point as $\sim\xi$. Both these properties are consistent with a behavior expected from an equilibrium scalar field theory in one spatial dimension where only Gaussian fluctuations are taken into account. Indeed, here, the hydrodynamic scaling and the smallness of the noise ensure that, within our approach, only Gaussian fluctuations appear at leading order in $1/L$, see also the discussion in Sec.~\ref{sec:consistency}.

It is also interesting to compare the behavior to that of the ABC model which also presents a singular correlation function when it phase separates (as computed in Ref.~\cite{bodineau_long_2008}). In contrast to what we find, this singularity is not related to a divergent length scale. This is related to the fact that the quasi-potential of the ABC model is always long-ranged~\cite{bodineau_long_2008}, as seen e.g.~at equal densities~\cite{evans_phase_1998,evans_phase_1998-1}. We return to the latter point below and a detailed comparison between the critical behavior in the present model and the phase transition in the ABC model is presented in Sec.~\ref{sec:finite-size-results} for periodic systems.

While all the correlation functions presented in Eqs.~(\ref{eq:corr})-(\ref{eq:cross}) depend on the two length scales discussed above, their behavior is, as expected, distinct.
Both $C_2^\rho$ and $C_2^m$ are even and continuous, but $C_2^\rho$ is monotonic and positive while $C_2^m$
is non-monotonic and changes sign as its argument is varied, see the inset of Fig.~\ref{Fig:twopoint}(b). Its extrema are located at $|x|=x^*\equiv 2\ell\xi\log\xi/(\xi-1)\sim \ell\log \xi$. On the other hand,
$C_2^{\rho, m}$
is an odd function and is discontinuous at the origin [see Fig.~\ref{Fig:twopoint}(c)].  These unusual properties can be best understood by rewriting Eq.~\eqref{eq:cross} in terms of the $+,-$ two-point functions 
\begin{equation}
C_2^{\rho,m}
=-\langle\rho_+(x)\rho_-(x')\rangle+\langle\rho_-(x)\rho_+(x')\rangle.
\end{equation}
With this decomposition, consider a fluctuation localized at position~$x$ with an enhanced $+$ particle density. It is quite clear that it will be positively correlated with an enhanced $-$ particle density fluctuation to the right of~$x$. The reason being that the $-$ particles to the right of $x$ would be impeded by the surge in $+$ particles to their left, causing them to accumulate. However, the effect on the particles to the left of $x$ is exactly the opposite. There, the self-propulsion of the $-$ particles is directed away from the local $+$ particles surge. These features are manifested by the correlation functions between the densities of either positive and negative particles as shown in Fig.~\ref{Fig:twopoint}(a).

Finally, we note that the Gaussian fluctuating hydrodynamic equations also allow us to determine the dynamical two-point functions. The results, most conveniently written in Fourier space, are given by:
\begin{eqnarray}
	\label{eq:c2_k}
	\tilde{C}_2^\rho
	&=& 
	\langle \delta \rho(k,\omega) \delta \rho(k',\omega') \rangle = \frac{2 \rho_0 (1 - \rho_0)}{ \ell L M(k, \omega)}  \left[ k^2 \omega^2 + \left\{k^2 + 2 + \frac{2 \xi^2 - 1}{\xi^2 (2 \rho_0 -1)}  \right\} k^2 (k^2+2) \right] \delta(k + k') \delta(\omega + \omega') , 
	\\
	\nonumber 
	\tilde{C}_2^m 
	&=& 
	\langle \delta m(k,\omega) \delta m(k',\omega') \rangle = \frac{2\rho_0}{\ell L M(k, \omega)} \left[ ( {\xi^{-2}} - 2) (1 - 2\rho_0) k^4 + (k^4 + \omega^2)(k^2 + 2)  \right] \delta(k + k') \delta(\omega + \omega'), 
	\\
	\nonumber
	\tilde{C}_2^{\rho, m} 
	&=& 
	\langle \delta \rho(k,\omega) \delta m(k',\omega') \rangle = \frac{2 i k \text{Pe} \rho_0 (1- \rho_0)}{ \ell L M(k, \omega)} \left[ 2 i k^2 \omega  (1 - \rho_0) -2 \rho_0 k^2 (k^2+2) \right]\delta(k + k') \delta(\omega + \omega'),
\end{eqnarray}
where $M(k, \omega) = [k^2 (k^2 + \xi^{-2}) - \omega^2]^2 + 4 (k^2+1)^2 \omega^2$. Most notably, by examining scaling properties of the correlation function we find that the dynamical scaling exponent for our model is given by $z=4$ on scales smaller than the correlation length and $z=2$ on larger scales, in agreement with a mean-field model B and consistent with the mean-field divergence of the static correlation function. Numerical results supporting this are presented below.

\subsection{Self-consistency Ginzburg type criterion}
\label{sec:consistency}
As detailed below in Sec.~\ref{sec:derivation-two-point} the correlation functions are obtained by linearizing the noisy hydrodynamic equations about a homogeneous profile. The divergence of the amplitude of the fluctuations exhibited by the two-point function~\eqref{eq:corr} as criticality is approached indicates that the assumption of small fluctuations may not be self-consistent near the critical point. To check this 
we employ a Ginzburg criterion and check if:
\begin{equation} \label{Eq:ginzburg}
{\langle \delta \phi (\xi)^2 \rangle} \ll 1,
\end{equation}
Here the average density fluctuation inside a correlation length is 
$\delta \phi (\xi) \equiv ({\xi} \ell)^{-1} \int_{\xi \ell} d x ~  \delta \rho(x)$, and we assume that the average density is of the order of 1. We note this analysis is equivalent to checking whether $\langle [\delta \rho (x) - \delta \rho (0)]^2 \rangle \ll 1$ holds for any value of $x$.
In the small $\ell$ limit, and using the Eqs.~\eqref{eq:c2_k} the condition reads
\begin{equation} \label{Eq:ginzburg_L}
\frac{\mathrm{Pe}^2 \xi }{\ell} {\rho_0^2 (1 - \rho_0)^2} \ll L
\,.
\end{equation}
Using the asymptotics \eqref{div} and omitting $\mathcal O\left(1\right)$ pre-factors, we find that the above condition can be rewritten as
\begin{equation}
\frac{1}{L\ell}\ll\sqrt{\text{Pe}_c-\text{Pe}} 
\:.\label{cond}
\end{equation}
The Ginzburg criterion implies that for any finite $\xi$, or $\text{Pe}_c-\text{Pe}$, a large enough $L$ can be chosen so that it is satisfied. This specifies how the thermodynamic limit has to be taken so that the calculations of the correlation functions, around the homogeneous profiles, are valid. Note that exactly at the critical point the calculation is not self-consistent.
While beyond the scope of this manuscript, it is interesting to ask if this is due to a breakdown of the macroscopic hydrodynamic description used here to describe the system or due to the contribution of non-linearities to the fluctuations.

\subsection{Free-energy for the density profile}
\label{sec:free-energy-density}
The steady state probability distribution of observing density and polarization profiles $\rho(x)$ and $m(x)$ takes a large-deviation form at large $L$ as 
$e^{-L \mathcal F[\rho(x),m(x)]}$ where $\mathcal F[\rho(x),m(x)]$ is an analog of a free-energy density. Our linearized theory
detailed in Sec.~\ref{sec:derivation-two-point}
allows us to probe the large-deviation function $\mathcal F$ up to Gaussian order.
In particular, the deviations from the most probable profiles, outside the binodal curve, take the form 
\begin{equation}
\label{free}\mathcal F[\rho(x),m(x)]=\frac{1}{2}\int dk\left[
\begin{array}{c}
\delta \tilde{\rho}(k) \\
\delta \tilde{m}(k)
\end{array}
\right]\mathbb{C}^{-1}_k
\left[
\begin{array}{c}
\delta \tilde{\rho}(-k) \\
\delta \tilde{m}(-k)
\end{array}
\right]+ o\big[(\delta\rho,\delta m)^2\big] \;,
\end{equation}
in Fourier space, where $\tilde{f}$ denotes the Fourier transform of $f$, $\mathbb{C}_k$ is a Hermitian matrix with diagonal elements given by~\eqref{Eq:Cor_rho_k} and~\eqref{Eq:Cor_m_k}, and the off-diagonal elements given by~\eqref{Eq:Cor_rho_m_k} and its complex conjugate \footnote{In \unexpanded{\eqref{free}} and \unexpanded{\eqref{free2}} we adopt the re-scaling used in Sec.~\ref{sec:derivation-two-point}. Also, note that the zeroth mode $k=0$ should be excluded from the integration above due to total particle number conservation}. Similarly, the free energy $\mathcal F[\rho(x)]$ corresponding to total density fluctuations is obtained by integrating over the magnetization fluctuations:
\begin{equation}
\mathcal F[\rho(x)]=\frac{1}{2}\int dk\frac{|\delta \tilde{\rho}(k)|^2}{C_2^{\rho}(k)}+\mathcal O(\delta\rho^4)\label{free2}
\end{equation}
with $C_2^{\rho}(k)$ given by~\eqref{Eq:Cor_rho_k}.
 
 The long-wavelength physics is captured by the small-$k$ expansion, where we find that $\mathcal F[\rho(x)]$ takes the form of the standard Landau theory near the critical point
\begin{equation}
\mathcal F[\rho(x)]\simeq\left(2\pi\right)^2\int_0^1 dx \left\{ \frac{\alpha}{\xi^2}\delta \rho^2+\beta\ell^2 \Big(\frac{d\delta\rho}{dx}\Big)^2 \right\}
,
\label{free_rho} 
\end{equation} 
with
\begin{equation}
\alpha=\frac{1}{2\rho_0\left(1-\rho_0\right)\left[2+\text{Pe}^2\left(1-\rho_0\right)\right]}\quad;\quad\beta=\frac{\text{Pe}^2\left(2+\xi^{-2}\right)}{2\left[2+\text{Pe}^2\left(1-\rho_0\right)\right]^2}.
\end{equation}
As expected, the ``mass'' term in Eq.~\eqref{free_rho} vanishes as the critical point is approached and $\xi\rightarrow \infty$.
Namely, the form of the large deviation function is that of a Gaussian field theory. Note, however, that if a standard Landau-Ginzburg free energy is rescaled so that $x \to x/L$ the coefficient of $(\frac{d\delta\rho}{dx})^2$ would scale as $1/L$ and vanishes in the large $L$ limit since the cost of a domain wall is finite. In the model we study, in contrast, the cost of a domain wall scales with $L$ for any non-vanishing $\ell$.

It is interesting to compare the result to those obtained previously for the ABC model which also exhibits phase separation in one dimension. There the large deviation function has been computed at equal densities~\cite{clincy2003phase}, close to the equal density point~\cite{clincy2003phase}, and for arbitrary densities but at Gaussian order~\cite{bodineau_long_2008}. In striking contrast, the large deviation function of the ABC model does not display similar criticality associated with a diverging correlation length (see also Sec.~\ref{sec:finite-size-results}  for a comparison of the divergent features of the correlation functions between the ABC model and the model studied here).

Next, we
detail how the results described above are obtained.

\section{Derivation of the fluctuating hydrodynamics}
\label{sec:deriv-fluct-hydr}

As stated above, the hydrodynamics~\eqref{eq:+}-\eqref{eq:-} describes the noiseless dynamics reached by the system when $L\to\infty$. However, at large but finite $L$, the actual dynamics display fluctuations around the deterministic hydrodynamic equation. The full fluctuating dynamics can be written as
\begin{eqnarray} \label{Eq:rhopm_derivation}
\partial_t\rho_{\pm} (x, t) =-\partial_xJ_{\pm} (x, t) \pm K (x, t).
\end{eqnarray} 
Here $J_{\pm}(x,t)$ are the conservative particles fluxes, and $K(x,t)$ is the total reaction rate contributed from tumbles where $+$ particles transform into $-$ particles and vice versa. Both terms are stochastic fields which fluctuate around their deterministic expressions
\begin{eqnarray} \label{Eq:derivation_K} 
\langle {J}_\pm (x,t) \rangle =-D\partial^2_x\rho_\pm  \pm \lambda\partial_x \left[\rho_\pm ( 1-\rho ) \right]\quad;\quad \langle {K (x,t) } \rangle =\gamma \left( \rho_-  -\rho_+ \right).
\end{eqnarray}
When $L$ is large, their \textit{typical} fluctuations are small, and by central limit theorem, these 
are given by Gaussian noise terms~\footnote{The account of large fluctuation will be presented in a future publication~\cite{forthcoming}.}
\begin{eqnarray}
J_{\pm} (x,t)= \langle {J}_\pm (x,t) \rangle +\sqrt{\frac{D}{L}}\eta_{\pm} (x,t) \quad;\quad K (x,t) = \langle {K} (x,t) \rangle +\sqrt{\frac{\gamma}{L}}\eta_K (x,t)
\end{eqnarray}
where the noise terms $\eta_{\pm} (x,t ),\eta_K (x,t )$ are uncorrelated over macroscopic time and length scales so that the covariance matrix is expected to be proportional to $\delta(x-x')\delta(t-t')$. In addition, the terms in the covariance matrix are expected to depend on the local densities. Finding the covariance matrix is achieved by building on a relation between the model and two other known lattice gas models: one specifies the conservative noise $\eta_\pm (x,t)$, while the other the non-conserving noise $\eta_K(x, t)$. 

\subsection{Conservative dynamics and the ABC model}
\label{sec:abcmodel}
To obtain the conservative noise, we set the tumbling rate of the model to zero.
In this case, the diffusive part of the model maps to a driven ABC model~\cite{clincy2003phase}. 
In fact, to compute the covariance of the flux noise terms, it is sufficient to consider the non-driven case where $\lambda=0$
(as generic in MFT, in our weak-drive limit, the noise is independent of the drive~\cite{bertini_macroscopic_2015}).
In this limit, the model coincides with an ABC model with symmetric dynamics, first presented in~\cite{evans_phase_1998} (corresponding to $q=1$ in their notations). Within the model, each site can be occupied by either an $A,B$, or $C$ particle. The dynamics are such that chiral pairs $AB,BC,CA$ exchange with rate $q$ (in arbitrary units) into $BA,CB,AC$ while the reverse exchanges occur with rate $q$. The model is known to exhibit phase separation for any finite $q<1$ \cite{evans_phase_1998,evans_phase_1998-1} and when the rate $q$ is scaled with the system size as $q=e^{-\beta/L}$ the phase transition occurs at a finite value of $\beta=\beta_c$ \cite{clincy2003phase}.  The mapping follows by identifying $+$, $-$ and vacant states with $A$, $B$, and $C$ particle types respectively. The fluctuating hydrodynamics for this model was studied in~\cite{bodineau_long_2008,bodineau_phase_2011}. The derivation there was based on yet another mapping, with each of the three species $A$, $B$, and $C$ behaving as a SSEP, a gas composed of particles interacting only through excluded volume interactions~\cite{derrida_microscopic_2011}. The corresponding flux noise terms have a covariance:
\begin{eqnarray}
\langle\eta_{\pm} (x,t)\eta_{\pm} (x',t' )\rangle=2\rho_{\pm} (1-\rho_{\pm})\,\delta (x-x')\delta (t-t')\ \ ;\ \ \langle\eta_+ (x,t)\eta_- (x',t' )\rangle=-2\rho_+\rho_-\delta (x-x')\delta (t-t').\label{eq:flux}
\end{eqnarray}
Transforming to the density and polarization fields $\rho$ and $m$, this yields the conservative part of the fluctuating hydrodynamics presented in Eqs.~(\ref{eq:rho})-(\ref{covm}).
Note that in~(\ref{eq:flux}) the same-species diagonal variance terms coincide with those of the SSEP, as one might anticipate. At zero bias $\lambda=0$, the different species are neutral to one another, i.e., the dynamics restricted to that of only one of the species coincides with that of freely evolving gas of SSEP particles.
Nevertheless, the cross-correlation between the fluxes of different species is non-vanishing and negative. This stems from the swapping of adjacent particles of different signs. Indeed, such moves contribute to oppositely oriented fluxes of the two particle types which leads to negative correlations. With this, we now turn to discuss the non-conserving part of the dynamics.

\subsection{Accounting for tumbles}
The reaction dynamics corresponding to tumbles can be accounted for within the fluctuating hydrodynamics of lattice gas models using several equivalent methods such as the Doi--Peliti formalism~\cite{doi_second_1976,doi_stochastic_1976,peliti_path_1985}, a large deviation formalism for reactive lattice gas~\cite{de_masi_reaction-diffusion_1986,jona-lasinio_large_1993,lefevre_dynamics_2007,tailleur_mapping_2008,bodineau_current_2010}, or for typical fluctuations using a van Kampen expansion~\cite{gardiner1985handbook}. Here, where the focus is on typical fluctuations, we use a simpler approach.

Specifically, we note that the coarse grained scaling allows us to treat the total tumble term $K(x,t) \equiv K_-(x,t) - K_+(x,t)$ as the difference of two uncorrelated Poisson processes $K_+(x,t)$ and $K_-(x,t)$ which specify the tumbling rates of $+$ and $-$ particles, respectively. According to Eq.~\eqref{Eq:rhopm_derivation}, ${\cal K}(x,t) \equiv \int_x^{x+\Delta x} L \mathrm{d} x' \int_t^{t+\Delta t} \mathrm{d} t' ~ K\left(x', t' \right)$ specifies the total number of particles that tumbled into the $+$ state subtracted by the number of particles that tumbled out of this state, inside a mesoscopic compartment of length $\Delta x$ 
during a time interval $\Delta t$. 
Since the densities evolve over macroscopic time and length scales, they can be approximated as constant inside $\Delta x$ during the time interval $\Delta t$ if we take $\Delta x$ such that  $L\Delta x\sim L^{\delta}\gg1$ with $1<\delta<0$. 
Hence, the random variable ${\cal K}(x,t)$ is simply the difference of two independent Poisson processes with rates $\gamma\rho_-(x,t)L\Delta x$ and $\gamma\rho_+(x,t)L\Delta x$ evolving over a time $\Delta t$. 

Next, using the fact that the Poisson distribution can be well-approximated by a Gaussian distribution in the large $L$ limit, we can specify the fluctuations by calculating the first two cumulants of ${\cal K}(x,t)$. These are easily found to be
\begin{eqnarray}
	\langle {\cal K}(x,t) \rangle &=& \gamma \left[\rho_- (x,t) -\rho_+ (x,t)\right]L\Delta x \Delta t \\
	\left\langle {\cal K}^2 (x,t) \right\rangle &=& \langle {\cal K}(x,t) \rangle^2 + \gamma \left[\rho_- (x,t) + \rho_+ (x,t)\right]L\Delta x \Delta t~\;,
\end{eqnarray}
which then implies that
\begin{equation} \label{Eq:K}
	K\left(x, t\right) =\gamma\left[\rho_- (x,t) -\rho_+ (x,t)\right] +\sqrt{\frac{\gamma}{L} \left[\rho_- (x,t) + \rho_+ (x,t) \right]} \, {\eta}(x,t)\:,
\end{equation}
where $\eta\left(x,t\right)$ is a zero mean and unit variance white Gaussian field. Defining $\eta_K(x,t) \equiv \sqrt{\rho} \eta(x,t)$ with the variance
\begin{equation}
\langle\eta_K (x,t)\eta_K (x',t')\rangle=\rho\,\delta (x-x')\delta (t-t'),
\label{trans}
\end{equation} 
leads to the announced result~\eqref{covm}.

Finally, we remark that this derivation also elucidates the validity of the Gaussian approximation. To see this, we evaluate the third cumulant of $\delta {\cal K} (x,t)$ which reads
\begin{equation}
		\langle \delta {\cal K}^3 (x,t) \rangle = \gamma \left[ \rho_+ (x, t) - \rho_- (x, t) \right] L \Delta x \Delta t~,
\end{equation}
and is not accounted for by Gaussian random noise. To incorporate this into $K(x,t)$ of Eq.~\eqref{Eq:derivation_K}, we need to add a term of order $L^{-2/3}$, which is sub-leading when compared to the Gaussian contribution which is of order $L^{-1/2}$. 
Note, that to account for large but rare fluctuations of $K(x,t)$, which go beyond the Gaussian approximation we present here and are described by large deviations, one needs to account for the non-Gaussian contributions to the statistics of $K$~\cite{forthcoming}.

\subsection{Superposition}

To combine both the results presented above we consider a coarse-grained description of the biased diffusion dynamics given by~\eqref{eq:flux} in the presence of the additional tumbling reactions. The former relies on the dynamics obeying ``local equilibrium'' described by a local product measure parameterized by the local particle densities~\cite{spohn_large_1991}. This local equilibrium property is in fact necessary for establishing the more basic deterministic description~\eqref{eq:+}-\eqref{eq:-} and was proven to hold in~\cite{erignoux_hydrodynamic_2018}. 
That is, as one might expect, introducing a slow tumbling reaction to the ABC dynamics does not violate the local equilibrium property that holds on mesoscopic scales.

Thus, we see that the two dynamics can be superimposed. Since tumbles occur independently of the particle diffusion, the noise terms of the two dynamics are uncorrelated. Under the appropriate change of variables to $\rho$ and $m$ variables, one arrives at Eqs.~\eqref{eq:rho}-\eqref{covm}.

\section{Derivation of the two-point correlation functions}
\label{sec:derivation-two-point}

\subsection{Static Correlation Functions}
\label{sec:static}

In this section, we derive the expressions for the different correlation functions shown above. Our main focus is the case $\ell \ll 1$ where two distinct phases coexist in the system within the binodal. However, in subsection \ref{sec:finite-size-results} we also consider the case when $\ell$ is not small. While in this case one cannot identify two distinct phases, the stability of the homogeneous phase can be studied and the correlation functions in it can be evaluated. 

In the hydrodynamic limit $L\gg1$, the two-point functions are found by solving for small fluctuations $\rho=\rho_0+\delta\rho/\sqrt{L}+\dots$ and $m=\delta m/\sqrt{L}+\dots$, around the steady-state homogeneous profiles. 
For convenience we rescale the variables so that $x \rightarrow \ell\, x$, $t \rightarrow \gamma\,t$. Then $-1/\ell < x < 1/\ell$ and for small $\ell$ we can use the continuum Fourier transforms
\begin{equation}
	\delta\tilde{m}\left(k,t \right)= \frac{1}{2 \pi} \int dx~ e^{-i kx }\delta m\left(x,t\right)
	\quad;\quad
	\delta\tilde{\rho}\left(k,t \right)= \frac{1}{2 \pi} \int dx~ e^{-ikx}\delta\rho\left(x,t\right)
\end{equation}
in Eqs.~(\ref{eq:rho}) and (\ref{eq:m}) with $k$ continuous. We then arrive at the linear system of equations
\begin{equation} \label{Eq:rho_m}
	\partial_t \left(
	\begin{array}{c}
		\delta \tilde{\rho}(k, t) \\
		\delta \tilde{m}(k, t)
	\end{array}
	 \right) = - \mathbb{M}_k
	 \left(
	 \begin{array}{c}
	 	\delta \tilde{\rho}(k, t) \\
	 	\delta \tilde{m}(k, t)
	 \end{array}
	 \right) + \mathbb{R}(k,t)~.
\end{equation}
Here, the matrix $\mathbb{M}_k$ and the random noise $\mathbb{R}(k, t)$ are
\begin{equation} \label{Eq:M_k}
	\mathbb{M}_k = \left(
	\begin{array}{cc}
		k^2 & i Ck  \\
		iE k  & k^2 + 2
	\end{array}
	\right)
	\quad;\quad
	\mathbb{R}(k, t) = \left(
	\begin{array}{c}
		Aik\,\tilde{\eta}_1 (k, t) \\
		\sqrt{k^2 + 2} B\,\tilde{\eta}_2 (k, t)
	\end{array}
	\right)~.
\end{equation}
with 
$C=\text{Pe}\left(1-\rho_0\right)$ and $E=\text{Pe}\left(1-2\rho_0\right)$, 
$A= \sqrt{2\rho_0\left(1-\rho_0\right)/(\ell L)}$,  $B= \sqrt{2\rho_0/(\ell L)}$, $C=\text{Pe}\left(1-\rho_0\right)$ and $E=\text{Pe}\left(1-2\rho_0\right)$, 
and where $\tilde{\eta}_{1,2}$ are Gaussian white noises with variance
\begin{equation}
	\left<\tilde{\eta}_i(k,t_1) \tilde{\eta}_j(k', t_2)\right> = \delta_{i,j} \delta(k+k')\delta(t_1 - t_2).
\end{equation}
The steady-state correlations can now be readily calculated using the solution of Eq.~\eqref{Eq:rho_m}
\begin{equation}
	\left(
	\begin{array}{c}
		\delta \tilde{\rho} (k, t) \\
		\delta \tilde{m} (k, t)
	\end{array} \right)
	= \int_{-\infty}^t \mathrm{d} t' ~ e^{-\mathbb{M}_k (t- t')} \mathbb{R} (k, t').
\end{equation}
Note that the zeroth mode $k=0$ remains identically zero at all times due to conservation of the total number of particles in the system. This implies that the equal time correlation function vanishes at $k=k'=0$. Accounting for this using a delta function whose magnitude is proportional to the finite inter-modes spacing $l$ gives
\begin{eqnarray}
	\label{Eq:Cor_rho_k} C_2^\rho (k, k') &=& \left\{ \frac{A^2 \left[ 4 + (6 + CE) k^2 + 2k^4 \right] + B^2 C^2 (2 + k^2)}{4(1 + k^2) (2 + CE + k^2)}- \frac{2A^2+B^2C^2}{2\left(2+CE\right)}l\delta\left(k\right)\right\} \delta (k + k'), \\
	\label{Eq:Cor_m_k} C_2^m (k, k') &=& \left[ \frac{A^2 E^2 k^2 + B^2 (2 + k^2) (2 + CE + 2 k^2)}{4(1 + k^2) (2 + CE + k^2)}- \frac{B^2}{2}l\delta\left(k\right) \right]\delta (k + k'), \\
	\label{Eq:Cor_rho_m_k} C_2^{\rho,m} (k, k') &=& \frac{i k (2 + k^2) (A^2 E - B^2 C)}{4(1 + k^2) (2 + CD + k^2)} \delta (k + k')\,.
\end{eqnarray}
Performing the inverse Fourier transform on Eqs.~(\ref{Eq:Cor_rho_k})-(\ref{Eq:Cor_rho_m_k}), one arrives at Eqs.~(\ref{eq:corr})-(\ref{eq:cross}). Notice that the finite $\ell$ correction introduced at the origin $k=k'=0$ via the additional delta functions, results in a constant offset of the real space two point functions Eqs.~(\ref{eq:corr}) and (\ref{Eq:c2_m}). This $\mathcal O\left(1\right)$ offset ensures that the integral over each of the real space coordinates vanishes, to leading order in $\ell$, as it should, due to total particle number conservation.

\subsection{Dynamic Correlation Functions}
\label{sec:dynamic}

For dynamic correlation functions, it is useful to use the Fourier transforms 
\begin{equation}
\delta\tilde{m}(k,\omega)=\frac{1}{(2\pi)^2}\int dxdt\: e^{-i(kx-\omega t)}\delta m(x,t)
\quad;\quad
\delta\tilde{\rho}(k,\omega)=\frac{1}{(2\pi)^2}\int dxdt\: e^{-i(kx-\omega t)}\delta\rho(x,t).
\end{equation}
Inserting these into Eqs.~(\ref{eq:rho}) and~(\ref{eq:m}), one obtains at leading order in $L$:
\begin{equation} \label{Eq:rho_m_omega}
	\left(
	\begin{array}{c}
		\delta \tilde{\rho} \\
		\delta \tilde{m}
	\end{array}
	\right) = \mathbb{M}_{k,\omega}^{-1} \mathbb{R} (k, \omega)~.
\end{equation}
with 
\begin{eqnarray}
	\mathbb{M}_{k,\omega} = \left(
	\begin{array}{cc}
		-i \omega + k^2 & i Ck  \\
		iE k  & - i \omega+  k^2 + 2
	\end{array}
	\right)
	\quad;\quad
	\mathbb{R}(k, \omega) = \left(
	\begin{array}{c}
		Aik\,\tilde{\eta}_1 (k, \omega) \\
		\sqrt{k^2 + 2} B\,\tilde{\eta}_2 (k, \omega)
	\end{array}
	\right).
\end{eqnarray}
Here, the random noises satisfy
\begin{equation}
\left<\tilde{\eta}_i(k,\omega)\tilde{\eta}_j(k',\omega')\right>=\delta_{ij} \delta(k+k')\delta(\omega+\omega')\,.
\end{equation}
Solving for $\delta\tilde{\rho}$ and $\delta\tilde{m}$ and averaging over noise realizations one finds the two-point correlation functions Eq.~\eqref{eq:c2_k}. 
As expected these expressions agree with Eqs.~(\ref{eq:corr})-(\ref{eq:cross}) after performing the inverse Fourier transform and taking the equal-time limit. 
By rescaling $x \to bx$ and $t \to b^zt$ and observing the scaling behavior of the dynamic correlation given in Eq.~\eqref{eq:c2_k} we identify the dynamic scaling exponent $z$ as follows. As $b$ becomes large, the correlation function converges to the following asymptotic form given as 
\begin{equation}
    \tilde{C}_2^\rho \simeq \frac{2 \rho_0 ( 1 - \rho_0)}{\ell L} \frac{k^2 \omega^2 + 2 b^{2z} k^2 \left\{ 2 + \frac{1 - 2 \xi^2}{(1 - 2 \rho_0) \xi^2} \right\}}{ 4 \omega^2 + \left[ b^{z-4} k^2(k^2 + b^2 \xi^{-2}) - b^{-z} \omega^2 \right]^2 } b^{z-1} \delta(k + k') \delta (\omega + \omega'),
\end{equation}
where the rescaling leads to $k \rightarrow b^{-1} k$ and $\omega \rightarrow b^{-z} \omega$. 
The correlation function shows different scaling behaviors depending on whether the length scale of interest is larger than or smaller than the correlation length $\xi$. For the length scales much larger than the correlation length ($k \ll \xi^{-1}$), the terms in the square brackets behave as $[b^{z-2} k^2 \xi^{-2}]^2$.
Taking large $b$, the correlation function becomes a scaling function of $\tilde{C}_2^\rho \propto k^{-2}[4 (\omega/k^2)^2 + \xi^{-4}]$ only if $z=2$, which implies the diffusive scaling. 
In contrast, if the length scale of interest is much smaller than the correlation length ($k \gg \xi^{-1}$), due to divergence of the correlation length nearby the critical point for instance, the terms in the square brackets converges to $[b^{z-4} k^4]^2$. In this case, the correlation function becomes $\tilde{C}_2^\rho \propto k^{-6}[4 (\omega/k^4)^2 + 1]$ when $z=4$.

\begin{figure} [t!]
	\includegraphics[width=0.4\linewidth]{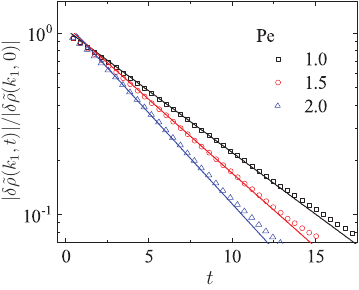}
	\caption{Relaxation dynamics of principal modes of the density fields starting from the initial configuration given in the square-wave form. Simulation results in symbols are compared with $\exp[ \lambda_- (k_1 ,\xi) t]$ shown in solid lines. Parameters: $L = 2^{11}$, $\gamma = 3 \times 10^2$, and $\rho_0 = 1/4$.
	}
	\label{Fig:relax}
\end{figure}

To check this numerically, it is more convenient to directly study the eigenvalues of $\mathbb{M}_k$. As Eq.~\eqref{Eq:rho_m} indicates, the eigenvalues determine the time scale for the relaxation of different modes. 
According to Eq.~\eqref{Eq:M_k}, the eigenvalues are given as 
\begin{equation} \label{Eq:lambda}
	\lambda_\pm = (k^2 + 1) \left[ 1 \pm \sqrt{1 - \frac{k^2 (k^2 + \xi^{-2})}{(k^2 + 1)^2}} \right].
\end{equation}
Here, we are interested in the long-wavelength dynamics governed by small-$k$ modes where
\begin{equation} \label{Eq:+-}
	\lambda_+ = 2 + \left( 2 - \frac{1}{2 \xi^2}  \right) k^2 - \frac{1}{8} \left( 2 - \frac{1}{\xi^2} \right)^2 k^4  + {\cal O}(k^6),  \quad \quad
	\lambda_- = \frac{1}{2 \xi^2} k^2 + \frac{1}{8} \left( 2 - \frac{1}{\xi^2} \right)^2 k^4 + {\cal O}(k^6)~.
\end{equation} 
Note that $\lambda_+ \rightarrow 2$ and $\lambda_- \rightarrow 0$ as $k \rightarrow 0$, so that the long-time dynamics of the fields are governed by $\lambda_-$ \footnote{The system also admits a different scaling regime at very small tumbling rates $\gamma\ll \lambda$ where the long wavelength dynamics is governed by sound waves corresponding to the eigenvalues $\lambda_{\pm}\simeq\pm i k\text{Pe}\sqrt{\left(1-2\rho_0\right)\left(1-\rho_0\right)}$. These decay over a time scale $\sim1/\gamma$ (in the dimension-full variables). This regime corresponds to $\text{Pe}\rightarrow\infty$ and it does not appear in our results \eqref{Eq:+-} as 
the $k\rightarrow0$ and $\text{Pe}\rightarrow \infty$ limits do not commute.
The wave velocity in this regime is real for $\rho_0<1/2$ where a homogeneous state is linearly stable but globally unstable. Since we only consider the behavior outside the binodal we do not consider it here.  }. In Fig.~\ref{Fig:relax}, we plot the relaxation dynamics of the principal mode when the system is initially prepared in a square-wave form with $\rho(x) = 1$ for $0 \leq x < \rho_0$ and $\rho(x) = 0$ for $\rho_0 < x<1$, and $m(x) = 0$. The simulation results, shown in symbols, display good agreement with the analytical relaxation $e^{-\lambda_- t}$ with $\lambda_-$ from Eq.~\eqref{Eq:+-}.

As the critical point is approached, the correlation length diverges and $\lambda_-$ becomes proportional to $k^4$ as shown by Eq.~\eqref{Eq:+-}, indicating the critical slowing down addressed with the correlation function above. Unfortunately, a direct numerical confirmation of the critical slowing down turned out to be beyond the reach of our numerics. Nearby the critical point, the Ginzburg criterion Eq.~\eqref{Eq:ginzburg_L} requires $L \gg \xi$ for the results to hold while the hydrodynamic time scale, required for the simulations, diverges as $L^2$. This makes it difficult to capture the long-time behavior of large systems in numerics using our numerical capabilities.  

\subsection{Correlation functions for finite  $\ell$}
\label{sec:finite-size-results}

In problems presenting a phase coexistence or a discontinuous phase transition,
finite systems can present different features than those in the thermodynamic limit. 
It is thus of interest to study how  finite-size effects manifest themselves in the system studied here. As we have shown, the finite system results also allow us to compare the critical behavior that we report here to that of the ABC model, which was studied in a finite system in~\cite{bodineau_long_2008}. 

To do this, it is instructive to first discuss the deterministic steady state of the system at finite $\ell$, as described by the noiseless hydrodynamics \eqref{eq:+}-\eqref{eq:-}. At finite $\ell$ the system no longer displays MIPS. Nevertheless, at large enough P\'eclet numbers, the homogeneous state becomes linearly unstable along a (modified) `spinodal' line~\cite{kourbane-houssene_exact_2018} which is found by analyzing the eigenvalues of the linearized dynamics Eq.~\eqref{Eq:lambda}.

Noting that at finite $\ell$ the discrete nature of the Fourier modes, $k=2\pi\ell n$ with $n$ a 
non-zero integer, has to be accounted for (the zeroth mode $k=0$ is stationary due to total particle number conservation) the first mode to become unstable is $k_1=2\pi\ell$. This happens when $\lambda_-$ changes sign at $k_1^2 + \xi^{-2} < 0$ (where $\xi$ in Eq.\eqref{xi} is purely imaginary), which sets the instability condition
\begin{equation} \label{Eq:instability_finite}
\text{Pe}^2 > \frac{2 + \left(2 \pi \ell \right)^2}{(1 - \rho_0) (2 \rho_0 -1)}.
\end{equation}
Note that at finite $\ell$, the instability sets in at larger P\'eclet values when compared to the small $\ell$ limit. In particular, there is no instability at $\text{Pe}=4$ and density $\rho=3/4$. However, as in the small $\ell$ limit, the homogeneous solution can become metastable before the `spinodal' curve \eqref{Eq:instability_finite} is reached. While at vanishing $\ell$ this happens along the binodal curve depicted in Fig.~\ref{fig:bin}~\cite{kourbane-houssene_exact_2018}, at finite $\ell$ at the moment there is no prescription for finding this line in a systematic manner. Therefore, in what follows, we are able to calculate the correlations about the flat state which may be, depending on the region in the phase diagram and outside the binodal, stable or meta-stable. Interestingly, even when the homogeneous state is metastable, for large $L$, it is long-lived. Since the fluctuations which drive the system out of the metastable state scale as $L^{-1/2}$, the mean time for the system to relax to the globally stable state is expected to grow exponentially with $L$ as $\exp(\alpha L)$ with $\alpha$ determined by the most probable path connecting the two states. The scaling with $L$ arises from the Gaussian cost of fluctuations, illustrated, for example, in the large-deviation function Eq.~\eqref{free}.

The derivation proceeds along the lines of Sec.~\ref{sec:derivation-two-point} where the continuous Fourier transform is replaced by a Fourier series for the fields $f(x) = \delta \rho(x), \delta m(x)$ in a system of finite length $1/\ell$:
\begin{eqnarray}
\tilde{f}_n = \ell\int_0^{1/\ell} \mathrm{d} x ~ f(x) e^{2 \pi i n\ell x}  
\quad ; \quad
f(x) = \sum_{n=-\infty}^{\infty}  \tilde{f}_n e^{- {2 \pi in\ell x}} ~.
\end{eqnarray}
\begin{figure} [t!]
	\includegraphics[width=0.4\linewidth]{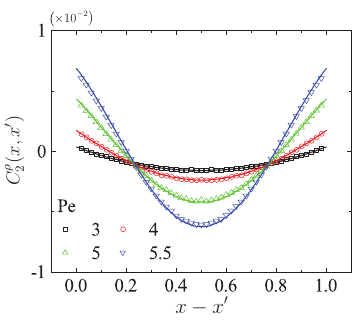}
	\caption{
		Two-point correlation functions displayed throughout the whole range $x - x' \in [0:1)$.  
		The average density is fixed at $\rho_0 = 3/4$ while $\mathrm{Pe}$ goes from 3 to 5.5. The simulation results are shown in symbols and compared to Eq.~\eqref{Eq:finite_rho2} shown in lines. Parameters: $L=2 \times 10^3$, $\gamma = 10^1$, and $D = 1$. Linear instability is expected at $\mathrm{Pe} = 6.90$ according to Eq.~\eqref{Eq:instability_finite}.}
	\label{Fig:finite}
\end{figure} 
With Eqs.~\eqref{eq:rho} and~\eqref{eq:m}, these give a discrete version of Eq.~\eqref{Eq:rho_m}
\begin{equation} \label{Eq:finite_rho_m}
\partial_t \left(
\begin{array}{c}
\delta \tilde{\rho}_n( t) \\
\delta \tilde{m}_n( t)
\end{array}
\right) = - \mathbb{M}_n
\left(
\begin{array}{c}
\delta \tilde{\rho}_n( t) \\
\delta \tilde{m}_n( t)
\end{array}
\right) + \mathbb{R}_n(t)~\;,
\end{equation}
where the matrix $\mathbb{M}_n$ and the noise $\mathbb{R}_n( t)$ are given by
\begin{equation} \label{Eq:finite_M_k}
\mathbb{M}_n = \left(
\begin{array}{cc}
k_n^2 & i Ck_n  \\
iE k_n  & k_n^2 + 2
\end{array}
\right)
\quad;\quad
\mathbb{R}_n( t) = \left(
\begin{array}{c}
Aik_n\,\tilde{\eta}_{1,n} ( t) \\
\sqrt{k_n^2 + 2} B\,\tilde{\eta}_{2,n} ( t)
\end{array}
\right)~.
\end{equation}
with $k_n = 2 \pi\ell n$ and $
\left<\tilde{\eta}_{i,n_1}\left(t_1\right) \tilde{\eta}_{j,n_2}\left(t_2\right)\right> = \ell\delta_{i,j} \delta_{n_1,-n_2}\delta\left(t_1-t_2\right)$. 
Solving the coupled equations and performing the inverse Fourier transform one finds, 

\begin{eqnarray} \label{Eq:finite_rho2}
\!\!\!\!
C_2^\rho (x, x') = \frac{\rho_0 ( 1 - \rho_0)}{L} \left[\delta (x - x') -1\right] +\frac{1}{L}
\frac{ \text{Pe}^2\xi^2\rho_0^2(1-\rho_0)^2}{2\ell(\xi^2-1)} \left[ f(|x-x'|) + g_\xi (|x-x'|)   \right]
\end{eqnarray}

where we defined 
\begin{equation}
f(|x-x'|) \equiv -\left[\frac{\cosh \left  (\frac{|x-x'|}{\ell} - \frac{1}{2\ell} \right)}{2\sinh \left( \frac{1}{2\ell} \right)}-\ell\right],\quad\text{and}
\end{equation}

\begin{numcases} 
{g_\xi (|x-x'|) \equiv} \frac{ 2 \xi^2-1}{\xi} \left[\frac{\cosh \left(  \frac{|x-x'|}{\ell\xi} - \frac{1}{2\ell\xi} \right)}{2\sinh \left(  \frac{1}{2 \ell\xi} \right)}-\ell\xi\right]&$0<\xi^{-2}$
\label{eq:fgxia}\\
\frac{1}{\ell}\left(\frac{1}{6}+|x-x'|^2-|x-x'|\right)&$ \xi^{-1} =0$
\label{eq:fgxib}\\
\frac{ 2 \xi^2-1}{|\xi|} \left[\frac{\cos \left(  \frac{|x-x'|}{\ell|\xi|} - \frac{1}{2\ell|\xi|} \right)}{2\sin \left(  \frac{1}{2 \ell|\xi|} \right)}-\ell|\xi|\right]&$-k_1^2<\xi^{-2}<0$ \;.
\label{eq:fgxic}
\end{numcases}
Here $|\xi|$ is the modulus of $\xi$, and $k_1=2\pi \ell$. Note that Eq.~\eqref{eq:fgxic} is the analytic continuation of Eq.~\eqref{eq:fgxia} obtained from the $0< \xi^{-2}$  regime to the $\xi^{-2} <0$ one where $\xi$ becomes purely imaginary. Also, notice that the integral of Eq.~\eqref{Eq:finite_rho2} over each of the coordinates $x,x'$ separately, vanishes due to total particle number conservation.
We compare the expressions for finite $\ell$ correlation to simulation results of the microscopic dynamics, all of which show excellent agreement as presented in Fig.~\ref{Fig:finite}. 

Eq.~\eqref{Eq:finite_rho2} can now be used to describe the finite $\ell$ rounding of the critical behavior predicted by Eq.~\eqref{eq:corr}. Consider first the limit where we take $\ell \to 0$ before approaching the critical point  $\left(\rho_c=3/4,\text{Pe}=4\right)$.
In this limit, we recover the infinite system result since Eq.~\eqref{Eq:finite_rho2} converges to Eq.~\eqref{eq:corr} and the region for Eq.~\eqref{eq:fgxic} shrinks to zero since $k_1 \to 0$. Here once $\xi$ becomes of the order of $1/\ell$, the small $\ell$ approximation breaks down and one must resort to the finite $\ell$ expression Eq.~\eqref{Eq:finite_rho2} where no singularity is present at $\text{Pe}=4$ as discussed above. For $\text{Pe}<4$ where $\xi^{-2}>0$ expression Eq.~\eqref{eq:fgxia} holds. At $\text{Pe}=4$ and $\rho_0=3/4$ the correlation function takes the form of the parabola Eq.~\eqref{eq:fgxib} in contrast to the $\ell \to 0$ where there is no valid analytical expression. Then for $\text{Pe}>4$ such that $-k_1^2<\xi^{-2}<0$ expression Eq.~\eqref{eq:fgxic} holds. These results are illustrated in Fig. \ref{Fig:finite}.

\begin{figure} [t!]
	\includegraphics[width=0.4\linewidth]{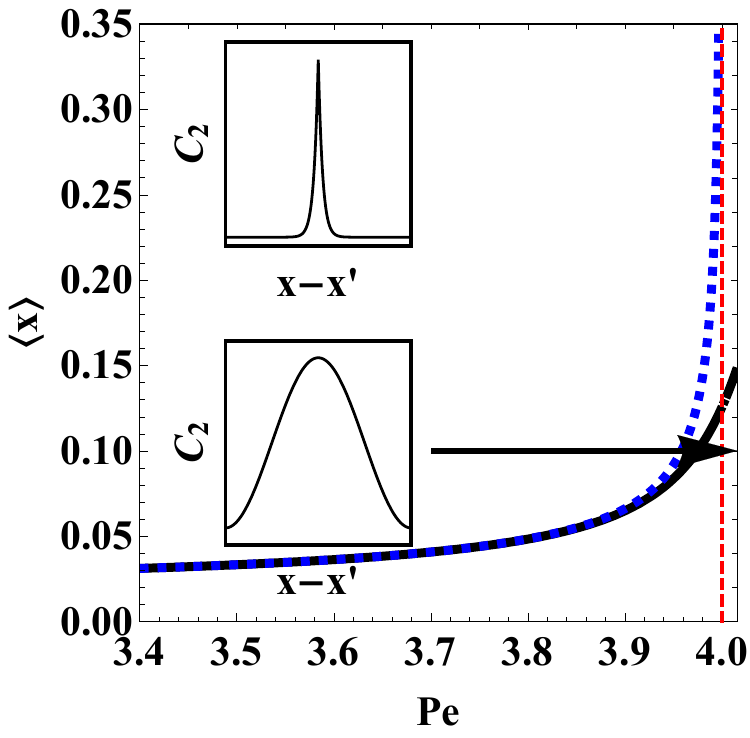}
	\caption{
		The first moment \eqref{first} at $\rho_0=\frac{3}{4}$ and $\ell=0.02$ as a function of the P\'eclet number. The dashed line corresponds to the vanishing $\ell$ expression \eqref{eq:corr}. The first moment grows from the value $\ell\ll1$ at $\text{Pe}=0$ until it spans the system and becomes $\mathcal O\left(1\right)$ around $\text{Pe}=4$ and up to the instability point at $\text{Pe}\approx 4.015$ [obtained from~\eqref{Eq:instability_finite}]. The two insets show the two point function \eqref{Eq:finite_rho2} well before $\text{Pe}=4$ (upper inset), and beyond that point (lower inset). }
	\label{length}
\end{figure}

Note that beyond $\text{Pe}=4$ for finite $\ell$ the correlation length spans the whole system size. To see this it is useful to evaluate the span of the two-point function using the first moment
\begin{equation}
\langle x\rangle\equiv \int_0^1dx~ \tilde{C}_2^{\rho}\left(x\right)x,\label{first}
\end{equation}
with
\begin{equation}
\tilde{C}_2^{\rho}\left(x\right)\equiv \frac{C_2^{\rho}\left(x\right)-\min_xC_2^{\rho}\left(x\right)}{\int_0^1dx\left[C_2^{\rho}\left(x\right)-\min_xC_2^{\rho}\left(x\right)\right]}
\end{equation}  
a normalized and shifted two point function so that serves as a proper distribution function. This measure grows from a small value which is of order $\ell$ for $\text{Pe}<4$ to the order of the system size at $\text{Pe}=4$ and remains at the order of the system size up to the linear instability point. This is illustrated in Fig.~\ref{length} and compared to the $\ell \to 0$ case where the span $\langle x \rangle$ diverges at $\text{Pe}=4$. 

Finally, the results above also allow one to compare the phase transition to that found in the version of the ABC model studied in~\cite{clincy2003phase}. In the ABC model, each site can be occupied by either an A, B, or C particle and the exchange dynamics lead to phase separation.
For average densities $r_{A,B,C}$ of the three species, and denoting $\Delta=1-2(r_A^2+r_B^2+r_C^2)$, it was shown that a continuous transition into a phase separated state occurs at $\beta=\beta_c\equiv 2\pi/\sqrt\Delta$~\cite{clincy2003phase}. Moreover, in the homogeneous phase the connected two-point correlation function is~\cite{bodineau_long_2008}:
\begin{equation}
\label{eq:ABCcorrelAAnonloc}
\langle\rho_A(x)\rho_A(x')\rangle_c
=
\frac 1L r_A(1-r_A)\big(\delta(x-x')-1\big) 
-\frac 3L r_A^2 r_B r_C
\bigg(
\frac{\cos \frac{\beta \sqrt\Delta (1-2x-2x')}{2}}{\sqrt \Delta \sin \frac{\beta \sqrt\Delta}{2}}
-
\frac{2}{\beta\Delta}
\bigg)
\:.
\end{equation}
Note that at the critical point the amplitude of the non-local contribution~(\ref{eq:ABCcorrelAAnonloc}) to the correlation function diverges due to its denominator, in the same way as is observed in Eq.~\eqref{eq:fgxic}.
However, this phase transition is not associated with a divergent correlation length, in contrast to the critical features of the active model studied in this paper.
Indeed, in the ABC model, Eq.~(\ref{eq:ABCcorrelAAnonloc}) shows that the correlation length spans the entire system also away from criticality, and does not grow as $\beta\to\beta_c$,
while in our case $\xi\ell$ does become formally divergent as $\text{Pe} \to \text{Pe}_{{c}}$ (i.e.~tends to the full system size).
In comparison to the model we study here, in the ABC model the domain wall width, which is known to behave as $1/|\ln q| \propto L/\beta$ is always of the order of the system size. In our case, the domain wall width is controlled by $\ell$ which can be held at any value as the critical point is approached. In particular, in the $\ell \to 0$ limit the domain wall vanishes. Then our system falls into the more conventional phase separation picture. In contrast, in the ABC model the domain wall is not an independent parameter and takes a large value at the critical point. In sum, in comparison to the ABC model, the model presents more standard features of critical phenomena.

\section{Discussion}
\label{sec:discussion}    

In this paper, we build upon a recent progress in deriving the exact noiseless hydrodynamics of active systems~\cite{erignoux_hydrodynamic_2018,kourbane-houssene_exact_2018}. We complement the noiseless description by accounting for fluctuations at the Gaussian level as given in \eqref{eq:rho}-\eqref{eq:cov0}. This description enabled us to compute the two-point correlation functions in the homogeneous phase, and to characterize exactly the critical behavior exhibited by the model. Such a treatment has thus far been missing, and provides a first conclusive analytical evidence that, at least for the model studied here, MIPS belongs to the mean-field Ising universality class when the infinite system size is taken properly. Our findings add to a number of numerical works~\cite{PhysRevLett.123.068002,Caballero_2018}, and phenomenological field theoretical studies~\cite{PhysRevLett.123.068002,maggi2021universality} which argue the same, even if the models studied previous are distinct in that the rates do not scale with the system size.
In addition, we considered the scaling of dynamic correlation functions and relaxation eigenvalues of long-wavelength excitations near the critical point and found that they exhibit critical slowing down with a mean-field model B exponent $z=4$ (although, as we discussed in Sec.~\ref{sec:dynamic}, such exponent was out of our numerical reach). It would be interesting to study correlation in the non-homogeneous phase where the length scale $\ell$ should have a role.

It is interesting to place the results in the context of general results on non-equilibrium systems which phase separate in one dimension (see for example~\cite{arndt1999spontaneous,lahiri1997steadily,lahiri2000strong,kafri2002criterion,kafri2003phase}). In those works non-diffusive systems were studied and a condition for the existence of phase separation, based on the dynamics of domain walls, was derived~\cite{kafri2002criterion}. Here the class of systems is rather different in that it is diffusive and is more in line with the diffusive ABC model introduced in~\cite{clincy2003phase}. While in the latter phase separation was a result of a chiral structure it would be interesting to understand how phase separation occurs in the active model studied here. In particular, it is not clear what features of the domain wall dynamics indicate whether a domain is small or large so that system can eventually phase separate. 
 
While the manuscript derives the fluctuations at the Gaussian level, in a forthcoming publication~\cite{forthcoming} we will present the complete account of large fluctuations beyond the Gaussian approximation within an adapted Macroscopic Fluctuation Theory. 

Acknowledgments --~We are grateful to A.~Frishman and J.~Tailleur for discussions and a careful reading of the manuscript. YK and SR are supported by an ISF grant and the National Research Foundation of Korea (2019R1A6A3A03033761).
VL acknowledges financial support from the ERC Starting Grant No.~680275 MALIG, the ANR-18-CE30-0028-01 Grant LABS and the ANR-15-CE40-0020-03 Grant LSD.

\bibliographystyle{apsrev4-1}
\bibliography{activ_gauss}

\end{document}